\documentclass[11pt,reprint,onecolumn,notitlepage,longbibliography,superscriptaddress]{revtex4-1}

\usepackage{graphicx}
\usepackage{amsmath} 
\usepackage{mathtools}
\usepackage{amsfonts}
\usepackage{bbm}
\usepackage{braket}
\usepackage{caption}
\usepackage{graphicx}
\usepackage{float}
\usepackage{caption}
\usepackage{subcaption}
\usepackage{url}
\usepackage{color}

\begin{document}

\title{Quantum computing methods for electronic states of the water molecule}

\author{Teng Bian}
\affiliation{Department of Physics, Purdue University, West Lafayette, IN, 47907 USA}

\author{Daniel Murphy}
\affiliation{School of Physics, Georgia Institute of Technology, Atlanta, Georgia, 30332-0400, USA}

\author{Rongxin Xia}
\affiliation{Department of Physics, Purdue University, West Lafayette, IN, 47907 USA}

\author{Ammar Daskin}
\affiliation{Department of Computer Engineering, Istanbul Medeniyet University, Kadikoy, Istanbul, Turkey}

\author{Sabre Kais}
\email{kais@purdue.edu}
\affiliation{Department of Physics, Purdue University, West Lafayette, IN, 47907 USA}
\affiliation{Department of Chemistry, Purdue University, West Laffayete, IN, 47907 USA}

\renewcommand{\thefootnote}{\fnsymbol{footnote}}

\footnotetext[1]{kais@purdue.edu}

\begin{abstract}

We compare recently proposed methods to compute the electronic state energies of the water molecule on a quantum computer. 
 The methods include the phase estimation algorithm based on Trotter decomposition, the phase estimation algorithm  based on the direct implementation of the Hamiltonian, direct measurement based on the implementation of the Hamiltonian and a specific variational quantum eigensolver, Pairwise VQE. After deriving the Hamiltonian using STO-3G basis, we first explain how each method works and then compare the simulation results in terms of gate complexity and the number of measurements for the ground state of the water molecule with different O-H bond lengths. 
Moreover, we present the analytical analyses of the error and the gate-complexity for each method. While the required number of qubits for each method is almost the same, the number of gates and the error vary a lot. In conclusion, among methods based on the phase estimation algorithm, the second order direct method provides the most efficient circuit implementations in terms of the gate complexity. With large scale quantum computation, the second order direct method seems to be better for large molecule systems. Moreover, Pairewise VQE serves the most practical method for near-term applications on the current available quantum computers. Finally the possibility of extending the calculation to excited states and resonances is discussed.\\
\\
\noindent
[Keywords]: Water molecule; Quantum computing methods; Electronic States; Resonances.

\end{abstract}

\maketitle

\section{Introduction}

The problem at the heart of computational chemistry is electronic structure calculation. This problem concerns calculating the properties of the stationary state describing many electrons interacting with some external potential and between each other via Coulomb repulsion. 
The ability to efficiently solve these problems for the cases of many body systems can have huge effects in pharmaceutical development, materials engineering, and all areas of chemistry. Quantum computing proposes the possibility to efficiently solve this problem for molecules with many more electrons than what can currently be simulated by classical computers\cite{kais_book}.

The ability to calculate properties of large quantum systems using precise control of some other quantum system was first proposed by Feynman\cite{feynman1982simulating}. He pointed out that if you have enough control over the states of some quantum system, you can create an analogy to some other quantum system. Using the example of spin in a lattice imitating many properties of bosons in quantum field theory, he conjectured that if you have enough individual quantum systems you could simulate any arbitrary quantum mechanical system. Simulation of the electronic structure Hamiltonian works very similar to this. Using the Jordan-Wigner or Bravyi-Kitaev transformation \cite{bravyi2002fermionic, setia2018bravyi, steudtner2017lowering} you can map an electronic structure Hamiltonian to a spin-type Hamiltonian which preserves energy eigenvalues \cite{xia2017electronic}. Evolution under this spin-type Hamiltonian, $e^{-iHt}$, can then be approximately simulated on quantum computers.

Quantum simulation provides a new and efficient way to calculate eigenenergies  of a given molecule. Classically the problem would have a computational cost which grows exponentially with the system size, $n$,  the number of orbital basis functions \cite{troyer2005computational}. 
However, based on the phase estimation algorithm \cite{lloyd1996universal,abrams1997simulation},  the molecular ground state energies can be calculated with gate depth $O(poly(n))$ \cite{aspuru2005simulated,whitfield2011simulation,wecker2014gate}. 
The quantum circuit for the Hamiltonian is generally approximated  through a Trotter-Suzuki decomposition. 
It is shown that the Hamiltonian dynamics can also be simulated through a truncated Taylor series \cite{berry2015simulating}. 
This method is generalized as quantum signal processing\cite{low2016hamiltonian}.
Babbush et al. \cite{babbush2017low} further shows that it is possible to reduce the gate depth of the circuit to $O(n)$ by using plane wave orbitals.
 Recently, a direct circuit implementation of the Hamiltonian within the phase estimation (Direct-PEA) is presented by  authors of paper \cite{daskin2017ancilla,daskin2017direct,daskin2018generalized}: the circuit designs are provided to the time evolution operator by using the truncated series such as $U = I-\frac{iH}{\kappa}$ and $U = tH+i(I-\frac{t^2H^2}{2})$, in which $\kappa$ and $t$ are parameters to restrict truncation error. 
 These unitary operators are much simpler to implement than those of a Trotter decomposition, and can be also used to calculate ground state energies of molecular Hamiltonians. 
 Another approach called variational quantum eigensolver (VQE) has been introduced by Aspuru-Guzik and coworkers\cite{peruzzo2014variational,mcclean2016theory}: This method combines classical and quantum algorithms together and significantly reduces the gate complexity at the cost of a large amount of measurements. 
 It has also been applied on real-world quantum computers to solve ground state energies of molecules such as: H$_2$, LiH and BeH$_2$ \cite{o2016scalable,kandala2017hardware}.

This paper explores all these above mentioned methods for calculating the ground state energies of the water molecule and presents a comparison study, in terms of both the accuracy and the gate complexity dependent on error. 
The next section explains the method by which the electronic Hamiltonian for water is calculated and the method by which to reduce the number of qubits required to simulate the transformed spin-type Hamiltonian. 
Then, Section III discusses five methods of electronic structure simulation on quantum computers:  the phase estimation using first order Trotter-Suzuki decomposed propagator (Trotter PEA), two direct implementations of the spin-type Hamiltonian (Direct PEA), a direct measurement and a specific variational quantum eigensolver method(Pairwise VQE). Section IV shows results for these methods with comparison to the exact energy from direct diagonalization of the spin-type Hamiltonian. It also gives qubit requirement and gate complexity for different methods asymptotically. Spin-type Hamiltonian for H$_2$O at equilibrium bond length is derived in Appendix A. Details of both error and complexity analyses are given in Appendix B and Appendix C.

\section{Hamiltonian Derivation}
In this section we provide details for calculating the spin-type Hamiltonian describing electronic structures of the water molecule using STO-3G basis set that will be used in later methods. This derivation can be generalized to an arbitrary molecular Hamiltonian. 

To obtain the Hamiltonian of the water molecule, we start by considering the 1s orbital of each hydrogen atom along with the 1s, 2s, 2p$_x$, 2p$_y$, 2p$_z$ orbitals for the oxygen atom. This leads to a total of 14 molecular orbitals considering spin. To make our simulations more efficient, the number of qubits is reduced by considering orbital energies and exploiting the symmetry of the system \cite{kandala2017hardware}. 

It can be initially assumed that the two molecular orbitals of largest energies are unoccupied. Consequently, the calculation of the Hamiltonian of the water molecule then only requires the consideration of 12 spin-orbitals. After second quantization, the Hamiltonian can be expressed as \cite{lanyon2010towards}: 
\begin{equation}
H= \sum_{i,j =1}^{12}h_{ij}a_i^\dagger a_j+\frac{1}{2}\sum_{i,j,k,l=1}^{12} h_{ijkl}a_i^\dagger a_j^\dagger a_ka_l.
\end{equation}
\noindent
Here $a_i^\dagger$ and $a_i$ are fermionic creation and annihilation operators, and $h_{i,j}$ and $h_{i,j,k,l}$ are one-body and two-body interaction coefficients. In this work the molecular orbitals are calculated from the Hartree-Fock method and represented by the STO-3G basis functions. The numerical integration obtaining the one and two electron integrals for molecular water is performed by the PyQuante package \cite{pyquante}. The expressions for these integrations are: \begin{equation}h_{ij} = \int{d\vec{r}_1\chi_i^*(\vec{r}_1)(-\frac{1}{2}\nabla_1^2-\sum\limits_{\sigma} \frac{Z_\sigma}{|\vec{r}_1-\vec{R}_\sigma|})\chi_j(\vec{r}_1)},\end{equation}
\begin{equation}h_{ijkl} = \int{d\vec{r}_1d\vec{r}_2\chi_{i}^*(\vec{r}_1)\chi_{j}^*(\vec{r}_2)\frac{1}{r_{12}}\chi_{k}(\vec{r}_2)\chi_{l}(\vec{r}_1)}.\end{equation}
\noindent
Here we have defined $\chi_i(\vec{r})$ as the $i^{th}$ spin-orbital, which is calculated from a spatial orbital obtained by the Hartree-Fock method and the electron spin states. $Z_\sigma$ is the $\sigma^{th}$ nuclear charge, $\vec{r_{i}}$ is the position of electron $i$, $r_{12}$ is the distance between the two points  $r_1$ and $r_2$, and $\vec{R}_\sigma$ is the position of $\sigma^{th}$ nucleus. 

We have ordered our spin-orbitals from 1 to 12 as follows: \{$1\uparrow,2\uparrow...,6\uparrow,1\downarrow,2\downarrow,...6\downarrow$\}, with first spin-up orbitals ordered from lowest to highest energy and continuing into spin-down orbitals ordered from lowest to highest energy. Now introduce an ad hoc set $F= \{1,2,7,8\}$ corresponding to the 4 lowest energy spin orbitals $\{1\uparrow,2\uparrow,1\downarrow,2\downarrow\}$. For the H$_2$O ground state, it can be assumed the spin orbitals in the set $F$ will be filled with electrons. The following one-body single electron interaction operators then become:
\begin{align}
    &a_{1}^\dagger a_{1} = 1, 
    \text{\ \ }a_{2}^\dagger a_{2} = 1,\text{\ \ }a_{7}^\dagger a_{7} = 1, 
    \text{\ \ }a_{8}^\dagger a_{8} = 1,\nonumber\\
    &a_{i}^\dagger a_{j} = 0,\text{\ if\ } i\neq j \text{\ , and\ } i \in F\text{\ or \ }j \in F.
\end{align}

\noindent
This assumption also allows us to simplify the two-electron interaction terms under certain conditions:
\begin{align}
a_i^\dagger a_j^\dagger a_ka_l =\left\{
\begin{aligned}
a_j^\dagger a_k & , & i=l,i\in F, \{j,k\} \notin F, \\
a_i^\dagger a_l & , & j=k,j\in F, \{i,l\} \notin F, \\
-a_j^\dagger a_l & , & i=k,i\in F, \{j,l\} \notin F, \\
-a_i^\dagger a_k & , & j=l,j\in F, \{i,k\} \notin F.
\end{aligned}
\right.    
\end{align}

\noindent
Moreover, this ability to neglect creation or annihilation operator with subscript from $\{1,2,7,8\}$, along with the ability to neglect two-body operators containing an odd number of modes in F, allows us to relabel our orbital set 1 to 8, corresponding to spin-orbitals: $\{3\uparrow,4\uparrow,5\uparrow,6\uparrow,3\downarrow,4\downarrow,5\downarrow,6\downarrow\}$

Using the parity basis and taking advantage of particle and spin conversation, the required qubit number can be further reduced\cite{seeley2012bravyi}. In the parity basis:
\begin{align}
    a_j^\dagger = X^\leftarrow_{j+1} \otimes \frac{1}{2}(X_j\otimes Z_{j-1}-iY_j),\\
    a_j = X^\leftarrow_{j+1} \otimes \frac{1}{2}(X_j\otimes Z_{j-1}+iY_j),
\end{align}
where
\begin{align}
    X^{\leftarrow}_{i} \equiv X_{n-1}\otimes X_{n-1} \otimes \cdots \otimes X_{i+1} \otimes X_i,\text{\ }n=8.
\end{align}

\noindent
This fermionic Hamiltonian can now be mapped to an 8-local Hamiltonian represented as a weighted sum of tensor products of Pauli matrix $\{I_i,X_i,Y_i,Z_i\}$, which almost preserves the ground state energy value. The new Hamiltonian in the electronic occupation number basis set can be mapped to the parity basis set as: 

\begin{align}
    \ket{f_1 f_2 ... f_8} \rightarrow \ket{q_1} \otimes \ket{q_2} \otimes ... \otimes \ket{q_8},
\end{align} 
\noindent
where 
\begin{align}
    q_i = \sum_{k=1}^i f_k \text{\ mod 2} \in \{0,1 \}.
\end{align}

\noindent
Here $f_k$ represents the number of electrons occupying the $k^{th}$ spin-orbital, and $q_k$ represents the sum of electron numbers from $1^{st}$ to $k^{th}$ spin-orbital. 

We can now assume that half of the left 6 electrons are spin-up and the other half are spin-down. If this is the case,  $\ket{q_4} = \ket{1}$, and $\ket{q_8} = \ket{0}$, which means only $Z_4,I_4,Z_8,I_8$ will apply on these states\cite{bravyi2017tapering}. Since $Z_4\ket{q_4} = -\ket{q_4}, Z_8\ket{q_8} = \ket{q_8}$, all $Z_4,Z_8$ can be substituted by $-I_4$ and $I_8$, with this assumption we can now reduce this problem to a 6-local Hamiltonian. 

\section{Methods of Simulation}
After the parity transformation and simplifications made above we now have a reduced 6-local Hamiltonian describing H$_2$O in the form: $H = \sum_{i=1}^{L}\alpha_ih_i$, where $\{\alpha_i\}$ is a set of coefficients, and $\{ h_i\}$ is a set of tensor products of the Pauli matrices $\{I_i,X_i,Y_i,Z_i\}$. Method A,B,C tries to evolve quantum system state by approximating the propagator $e^{-iHt}$, and then extract the ground state energy from the phase. Method D implements the Hamiltonian, H, directly into quantum circuit, and evaluate ground state energy by multiple measurements. Method E produces the ground state energies by iterations.

\subsection{Trotter Phase Estimate Algorithm (Trotter-PEA)}
 For each term in a Hamiltonian, $H$, the propagator, $e^{-i\alpha_i h_it}$, can be easily constructed in a circuit. However, since most of the time the set of $h_i$ do not commute, the propagator cannot be implemented term by term: i.e., $e^{-iHt}\ne\prod_{i=1}^{L}e^{-i\alpha_ih_it}$. The first order Trotter-Suzuki decomposition \cite{trotter1959product,suzuki1976generalized,dhand2014stability} provides an easy way to decompose a propagator for the spin-type Hamiltonian given as a sum of non-commuting terms into a product of each non commuting term exponentiated for a small time $t$:
 \begin{align}
     U = \prod_{i=1}^{L}e^{-i\alpha_ih_i t} = e^{-iHt} + O(A^2t^2).
 \end{align}
Here $A = \sum_{i=1}^L |\alpha_i|$, and we have an error of order $O(A^2t^2)$. Here we don't consider time slicing as the original Trotter-Suzuki decompositoin does, as $t$ can be adjusted to be as small as necessary for error control. This method requires only multi-qubit rotations, and therefore $U$ can be implemented easily on a state register. 
 
 After $U$ is obtained PEA can be applied to extract the phase. We can use extra ancilla qubits to achieve wanted accuracy by iterative measurements\cite{kitaev1995quantum,dobvsivcek2007arbitrary,aspuru2005simulated}. We call this PEA based on first order Trotter-Suzuki decomposition Trotter-PEA. 
 
\begin{figure}[H]
 \centering
\includegraphics[]{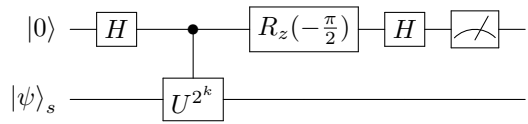}
\caption{Forward iterative PEA circuit with initial state $\ket{0}\ket{\psi}_s$. Here $\ket{\psi}_s$ is the ground state of the Hamiltonian, $H$ is the Hadamard gate, $U$ is the approximate propagator and $R_z(-\frac{\pi}{2})$ is a Z rotation gate. \label{Fig1}}
\end{figure}

 Higher order Trotter-Suzuki decompositions are also available, however they have more complicated formulations, especially for order higher than 2. Here we only discuss first order case for simplicity. For simulation, a forward iterative PEA \cite{daskin2018generalized}-which estimates the phase starting from the most significant bit-can be used to save more time. The circuit for the forward iterative PEA is shown in FIG. \ref{Fig1} which needs only 1 qubit for measurement.
 
\noindent 
Then the generated state before the measurement is:
\begin{align}
    \frac{1+e^{i2\pi(0.\phi_{k+1}\phi_{k+2}...-0.01)}}{2}e^{i\frac{\pi}{4}}\ket{0}\ket{\psi}_s + \frac{1-e^{i2\pi(0.\phi_{k+1}\phi_{k+2}...-0.01)}}{2}e^{i\frac{\pi}{4}}\ket{1}\ket{\psi}_s,
\end{align}

\noindent
Note decimals above are in binary. It can be checked if the measurement qubit has a greater probability of output 1, $\phi_{k+1} = 1$, otherwise $\phi_{k+1} = 0$. Then the ground state energy can be calculated as $E = -2\pi \times 0.\phi_1\phi_2\phi_3...$.

\subsection{Direct Implementation of Hamiltonian in First Order (Direct-PEA ($1^{st}$ order))}
It was proposed\cite{daskin2017direct} that given a Hamiltonian $H$ and large $\kappa$ we can construct an approximated unitary operator $U$ such that: 
\begin{align}
    U=I-i\frac{H}{\kappa}, \text{\ } \kappa \gg \sum_{i=1}^L |\alpha_i| \geq ||H||.
\end{align}

\noindent
If $\ket{\psi}_s$ is an eigenvector of $H$ and $E$ is the corresponding eigenvalue, then:

\begin{align}
    U\ket{\psi}_s = \left(I-i\frac{H}{\kappa}\right)\ket{\psi}_s \approx e^{-i\frac{H}{\kappa}} \ket{\psi}_s = e^{-i\frac{E}{\kappa}} \ket{\psi}_s.
\end{align}
 
 \noindent
 The eigenvalue of $\ket{\psi}_s$ would be encoded directly in the approximate phase. This is the motivation behind directly implementing the Hamiltonian in quantum simulation.\\

\noindent
To implement this non-unitary matrix $U$, we can enlarge the state space and construct a unitary operator $U_r$\cite{berry2015simulating}. Rewrite U as:
\begin{align}
U = I - \frac{i}{\kappa} \sum_{j=1}^L \alpha_j h_j = \sum_{j=0}^{L} \beta_j V_j,
\end{align}
in which $\beta_j \geq 0$ and $V_j$ is unitary. By introducing a $m$-qubit ancilla register, where $m = \lceil \log_2 L \rceil$, we can construct a multi-control gate, $V$, such that:
\begin{align}
    V\ket{j}_a\ket{\psi}_s=\ket{j}_a V_j \ket{\psi}_s.
    \label{gateV}
\end{align}

\noindent
Define $\beta_j = 0$ when $ L < j \leq 2^m$ and $B$ as a unitary operator that acts on ancilla qubits as:
\begin{align}
B\ket{0}_a=\frac{1}{\sqrt{s}}\sum_{j=0}^{2^m}\sqrt{\beta_j}\ket{j}_a,\  s=\sum_{j=0}^{2^m}\beta_j.
\label{gateB}
\end{align}

\noindent
Define $U_r$ and $\Pi$ such that:
\begin{align}
    &U_r = (B^\dagger \otimes I^{\otimes n}) V (B \otimes I^{\otimes n}),\\
    &\Pi = \ket{0}_a \bra{0}_a \otimes I^{\otimes n}.
\end{align}

\begin{figure}[H]
 \centering
\includegraphics[]{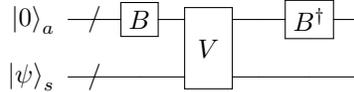}
    \caption{Gate $U_{r}$ in Direct PEA circuit, gates $V$ and $B$ are shown in Eq. (\ref{gateV}) and Eq. (\ref{gateB})}
    \label{ur.png}
\end{figure}

\noindent
Apply $U_r$ on input state $\ket{0}_a \ket{\psi}_s$:
\begin{align}
    U_r \ket{0}_a \ket{\psi}_s &= (B^\dagger \otimes I^{\otimes n}) V (B \otimes I^{\otimes n}) \ket{0}_a \ket{\psi}_s \nonumber\\
    &= (B^{\dagger} \otimes I^{\otimes n}) V \frac{1}{\sqrt{s}}\sum_{j=0}^{2^m}\sqrt{\beta_j}\ket{j}_a \ket{\psi}_s \nonumber\\
    &= (B^{\dagger} \otimes I^{\otimes n}) \frac{1}{\sqrt{s}}\sum_{j=0}^{2^m}\sqrt{\beta_j}\ket{j}_a V_j \ket{\psi}_s \nonumber\\
    &= \Pi (B^{\dagger} \otimes I^{\otimes n}) \frac{1}{\sqrt{s}}\sum_{j=0}^{2^m}\sqrt{\beta_j}\ket{j}_a V_j \ket{\psi}_s+ (I^{\otimes m+n} - \Pi) (B^{\dagger} \otimes I^{\otimes n}) \frac{1}{\sqrt{s}}\sum_{j=0}^{2^m}\sqrt{\beta_j}\ket{j}_a V_j \ket{\psi}_s \nonumber\\
    &= (B\ket{0}_a)^\dagger \frac{1}{\sqrt{s}}\sum_{j=0}^{2^m}\sqrt{\beta_j}\ket{j}_a V_j \ket{\psi}_s + \sum_{j=1}^{j=2^m}\ket{j}_a \ket{u_j}_s \nonumber\\
    &= \frac{1}{s} \ket{0}_a U \ket{\psi}_s + \ket{\Phi^\perp_1},
\end{align}

\noindent
where $\ket{\Phi^\perp_1}$ is orthogonal to $\ket{0}_a \ket{\psi}_s$.  Then the approximated unitary operator $U$ is implemented by unitary operator $U_r$, which can be seen in FIG. \ref{ur.png}.

\noindent
Since $\kappa \gg ||H|| \geq E$, energy of eigenstate $\ket{\psi}_s$ is successfully implemented in phase:
\begin{align}
    U_r \ket{0}_a \ket{\psi}_s &= \frac{1-i\frac{E}{\kappa}}{s} \ket{0}_a \ket{\psi}_s + \ket{\Phi^\perp_1} \nonumber\\
    &= \frac{\sqrt{1+\frac{E^2}{\kappa^2}}}{s} e^{-i\tan^{-1} \frac{E}{\kappa}} \ket{0}_a \ket{\psi}_s + \ket{\Phi^\perp_1} \nonumber\\
    &= pe^{-i\tan^{-1}\frac{E}{\kappa}} \ket{0}_a \ket{\psi}_s + \sqrt{1-p^2} \ket{\Phi^\perp}.
\end{align}
Here $p$ is defined by $\frac{\sqrt{1+\frac{E^2}{\kappa^2}}}{s}$, and $\ket{\Phi^\perp}$ is normalized.\\

This $U_r$ gate would then be used for PEA or iterative PEA process. For an accurate output, $p$ is required to be as close to 1 as possible. Using oblivious amplitude amplification\cite{berry2017exponential}, we can amplify that probability without affecting phase. Define the operator $U_0 =2\ket{0}_a \bra{0}_a-I^{\otimes m}$ and rotational operator: 
\begin{align}
    Q =U_r (U_0 \otimes I^{\otimes n}) U_r^\dagger (U_0 \otimes I^{\otimes n}).    
\end{align}
\noindent
Iterating this operator $N$ times, we can achieve $U_q = Q^NU_r$ which brings $p$ close to $1$ by performing rotations within the space $span\{\ket{0}_a \ket{\psi}_s, \ket{\Phi^\perp}\}$. The details are in Supplementary Materials. Take the same circuit and the same procedure in Trotter-PEA, except replacing $U$ by $U_q$, we are able to get ground state energy of water molecule.

\subsection{Direct Implementation of Hamiltonian in Second Order (Direct-PEA ($2^{nd}$ order))}
Propagator $e^{-iHt}$ can also be approximated up to second order\cite{daskin2018generalized}:
\begin{align}
    &U = I-iHt-\frac{H^2t^2}{2} = e^{-iHt} + O((At)^3).
\end{align}

\noindent
When $At$ is very small, $U$ would be a good approximation. Since $U$ is nonunitary, we have to construct a unitary operator $U_{r2}$ to implement it into a quantum circuit. With $U_r$ in method B, $B_2$ defined with the property: 
\begin{align}
    B_2 \ket{00} = \frac{\sqrt{t}\ket{00} + \ket{01} + \frac{t}{\sqrt{2}}\ket{10}}{\sqrt{1+t+\frac{t^2}{2}}},
    \label{gateB2}
\end{align}
and gate $P$ constructed as:

\begin{equation}
    P =
\begin{bmatrix}
    I^{\otimes n}& 0 & 0 & 0\\
    0 & 0 & I^{\otimes m} & 0\\
    0 & I^{\otimes m} & 0 & 0\\
    0 & 0 & 0 & I^{\otimes n}
    \label{gateP}
\end{bmatrix}.
\end{equation}

\noindent
We can construct $U_{r2}$ as in FIG \ref{ur2.png}:
 \begin{figure}[H]
 \centering
\includegraphics[]{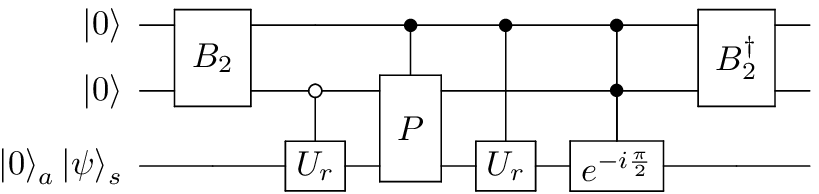}
    \caption{Gate $U_{r2}$ in Second order Direct PEA circuit, with $B_2$ and $P$ defined in Eq. (\ref{gateB2}) and Eq. (\ref{gateP})}
    \label{ur2.png}
\end{figure}

\noindent
which satisfies:
\begin{align}
    U_{r2}\ket{00}\ket{0}_a\ket{\psi}_s &=\frac{1-i\frac{Et}{A}+\frac{E^2t^2}{2A^2}}{1+t+\frac{t^2}{2}} \ket{00}\ket{0}_a\ket{\psi}_s + \sum_{j=1}^{2^{m+2}}\ket{j}\ket{v_j}_{s} \nonumber\\
    &= \frac{\sqrt{1+\frac{E^4t^4}{4A^2}}}{1+t+\frac{t^2}{2}} e^{-i\tan^{-1}{\frac{\frac{Et}{A}}{1+\frac{E^2t^2}{2A}}}} \ket{00}\ket{0}_a\ket{\psi}_s + \ket{\Psi_1^\perp}.
\end{align}

\noindent
In the formula, $A = \sum_{i=1}^{2^m-1} \beta_i = \sum_{i=1}^L |\alpha_i| \geq |E|$, and $\ket{\Psi_1^{\perp}}$ is perpendicular to $\ket{00}\ket{0}_a \ket{\psi}_s$. Just as in last section, we can rotate the final state to make the proportion of $\ket{00}\ket{0}_a\ket{\psi}_s$ as close to 1 as possible. Then we can apply PEA or iterative PEA to get the phase, $-\tan^{-1}{\frac{\frac{Et}{A}}{1+\frac{E^2t^2}{2A}}}$, which leads to ground state energy corresponding to ground state $\ket{\psi}_s$.

\subsection{Direct Measurement of Hamiltonian}
Another way to calculate the ground state energy is by direct measurement after implementing a given Hamiltonian as a circuit. Since Direct-PEA ($1^{st}$ order) method has already introduced a way to implement non-unitary matrix $U$ into circuit, Hamiltonian implementation is straightforward. We can just replace $U$ in method B by $U' = H = \sum_{j=1}^{L} \alpha_j h_j$, and obtain $U_r'$ such that:
\begin{align}
    U_r'\ket{0}_a\ket{\psi}_s &= \frac{1}{s'}\ket{0}_a U_r' \ket{\psi}_s + \ket{\Phi_1^{'\perp}} \nonumber\\
    &= \frac{E}{A} \ket{0}_a \ket{\psi}_s + \ket{\Phi_1^{'\perp}}.
\end{align}
\noindent
By measuring ancilla qubits multiple times, we can get the energy of the ground state $\ket{\psi}_s$ by multiplying $A$ by the square root of probability of getting all 0s.

This method can also be used for non-hermitian Hamiltonians. If now the eigenvalue for $\ket{\psi}_s$ is a complex number $E = |E|e^{i\theta}$, by replacing $U$ by $U' = H$ in method $B$, we would have:
\begin{align}
    U_r'\ket{0}_a\ket{\psi}_s = \frac{|E|e^{i\theta}}{A} \ket{0}_a \ket{\psi}_s + \ket{\Phi_1^{'\perp}},
\end{align}
and can obtain $|E|$ through measurements. Then by replacing $U$ by $U'' = \frac{|E|}{A}I + H$ in method $B$, we would have:
\begin{align}
     U_r''\ket{0}_a\ket{\psi}_s = \frac{|E|}{A}(1+e^{i\theta}) \ket{0}_a \ket{\psi}_s + \ket{\Phi_1^{'\perp}},
\end{align}
and can measure the absolute value of $\frac{|E|}{A} (1+e^{i\theta})$, which is $2\frac{|E|}{A} \cos \theta$. This helps determine the phase of a complex eigenenergy.

\subsection{Variational Quantum Eigensolver}
\begin{figure}[H]
\centering
\small
\includegraphics[]{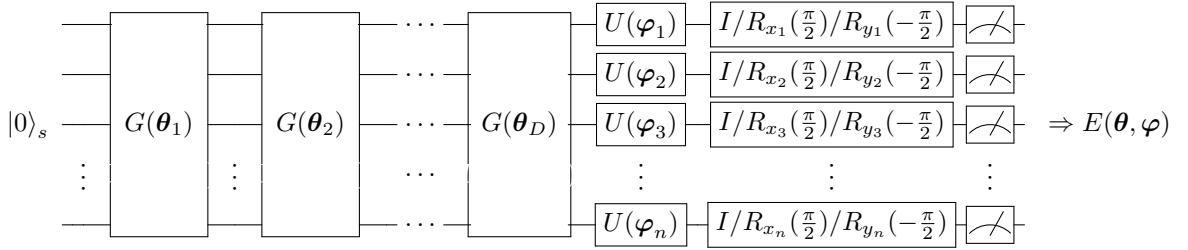}
    \caption{Circuit for state preparation and corresponding energy evaluation. $G({\boldsymbol\theta}_i)$ is entangling gate, in this paper we are taking the gate like FIG. \ref{vqe_v5.png}.  $U(\boldsymbol{\varphi}_k)$ is an arbitrary single-qubit rotation and is equal to $R_{z}(\varphi_{k,1})R_{x}(\varphi_{k,2})R_{z}(\theta_{k,3})$  with parameters $\varphi_{k,1}$,$\varphi_{k,2}$ and $\varphi_{k,3}$ that can be manipulated. By increasing the number of layers, $d$,  of our circuit, we are able to produce more complex states.}
    \label{vqe1.png}
\end{figure}

Recently the variational quantum eigensolver method has been put forward by Aspuru-guzik and coworkers to calculate the ground state energies\cite{peruzzo2014variational,mcclean2016theory,o2016scalable,kandala2017hardware,gilyen2017optimizing}, which is a hybrid method of classical and quantum computation. According to this method, an adjustable quantum circuit is constructed at first to generate a state of the system. This state is then used to calculate the corresponding energy under the system's Hamiltonian. Then by a classical optimization algorithm, like Nelder-Mead method, parameters in circuit can be adjusted and the generated state will be updated. Finally, the minimal energy will be obtained. The detailed circuit for the quantum part of our algorithm is shown in FIG.\ref{vqe1.png}. To make the expression more clear, we represent parameters in vector form, as follows: $\boldsymbol{\theta} = ( \boldsymbol{\theta_1}, \boldsymbol{\theta_2}..., \boldsymbol{\theta_D} ) $, $\boldsymbol{\theta_i} = (\boldsymbol{\theta_{i,0}}, \boldsymbol{\theta_{i,1}}..., \boldsymbol{\theta_{i,11})} $, $\boldsymbol{\theta_{i,j}} = (\theta_{i,j,1}, \theta_{i,j,2}, \theta_{i,j,3}, )$, $\boldsymbol{\varphi} = ( \boldsymbol{\varphi_1}, \boldsymbol{\varphi_2}..., \boldsymbol{\varphi_n} )$, $\boldsymbol{\varphi_k} = (\varphi_{k,1}, \varphi_{k,2}, \varphi_{k,3})$.

\begin{figure}[H]
\centering
\footnotesize
\includegraphics[]{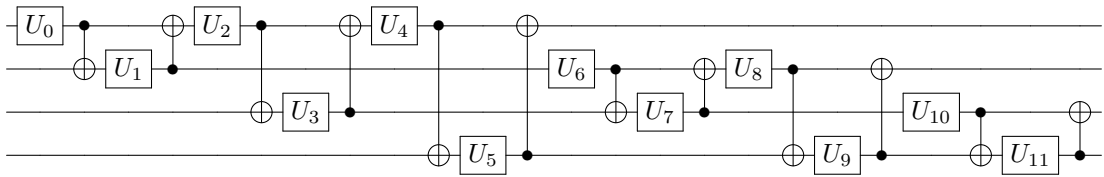}
    \caption{Example entangling circuit $G(\boldsymbol\theta_i)$ for 4-qubit system. There are 12 arbitrary single-qubit gates $U_j$, a simplified written way for $U(\boldsymbol{\theta}_{i,j})$, which is $R_{z}(\theta_{i,j,1})R_{x}(\theta_{i,j,2})R_{z}(\theta_{i,j,3})$ with parameters $\theta_{i,j,1}$,$\theta_{i,j,2}$ and $\theta_{i,j,3}$ that can be manipulated. Each 2 qubits are entangled sequentially. Entangling gate $G(\boldsymbol\theta_i)$ for $n$-qubit system is similar to this gate, but then it has $n(n-1)$ arbitrary single-qubit gates and $\boldsymbol{\theta}_i$ has $3n(n-1)$ parameters.}
    \label{vqe_v5.png}
\end{figure}
  
\noindent
  We are using $d$ layers of gate $G({\boldsymbol{\theta}}_i)$ in FIG. \ref{vqe1.png} to entangle all qubits together. Here we introduce a hardware-efficient $G({\boldsymbol{\theta}}_i)$, and we call this method Pairwise VQE. The example gate of $G({\boldsymbol{\theta}}_i)$ for 4 qubits is shown in FIG. \ref{vqe_v5.png}. The entangling gate for 6-qubit system H$_2$O is similar: every 2 qubits are modified by single-qubit gates and entangled by $CNOT$ gate. By selecting initial value of all $\boldsymbol\theta_i$ and $\boldsymbol\varphi_k$, system state can be prepared by $d$ layers $G({\boldsymbol{\theta}}_i)$ gates and arbitrary single gates $U({\boldsymbol{\varphi}}_j)$. Then average value of each term in Hamiltonian $H$, $\langle h_j \rangle$ , can be evaluated by measuring qubits many times after going through gates like $I$ or $R_{x_j}(\frac{\pi}{2})$ or $R_{y_j}(-\frac{\pi}{2})$. For example, if $h_j = I_0X_1Y_2Z_3$, then 
  \begin{align*}
  \langle h_j \rangle &= \langle I_0X_1Y_2Z_3\rangle_{\psi} = \bra{\psi} I_0X_1Y_2Z_3 \ket{\psi}\\
  &= (\bra{\psi} R_{y_1}(\frac{\pi}{2})R_{x_2}(-\frac{\pi}{2})) I_0 (R_{y_1}(-\frac{\pi}{2})X_1R_{y_1}(\frac{\pi}{2}))  (R_{x_2}(\frac{\pi}{2})Y_2R_{x_2}(-\frac{\pi}{2})) Z_3 (R_{y_1}(-\frac{\pi}{2})R_{x_2}(\frac{\pi}{2})\ket{\psi})\\
  &= \langle I_0 Z_1 Z_2 Z_3\rangle_{\psi'} \text{\quad, where} \ket{\psi'} = R_{y_1}(-\frac{\pi}{2})R_{x_2}(\frac{\pi}{2})\ket{\psi},
  \end{align*}
  So we can let the quantum state after $U(\boldsymbol{\varphi_j})$ go through gates $R_{y_1}(-\frac{\pi}{2})$ and $R_{x_2}(\frac{\pi}{2})$ and then measure the result state multiple times to get $\langle h_j \rangle$. The energy corresponding to the state can be obtained by $ \langle H \rangle (\boldsymbol{\theta},\boldsymbol{\varphi}) = \sum_{j=1}^{L} \alpha_j \langle h_j\rangle (\boldsymbol{\theta},\boldsymbol\varphi)$. Then $\boldsymbol{\theta}$ and $\boldsymbol{\varphi}$ can be updated by classical optimization method and $\langle H \rangle(\boldsymbol\theta, \boldsymbol\varphi)$ can reach the minimal step by step.
  

\section{Results and Method Comparison}
The Hamiltonian of the water molecule is calculated for O-H bond lengths ranging from 0.5 a.u. to 2.9 a.u., using the methods introduced in Section II. This Hamiltonian is used in all five of the methods discussed within this paper. For the methods  A-D, the input state of system is the ground state of the H$_2$O molecule. For each of these methods, the resulting ground state energy curve can be calculated to arbitrary accuracy (for details of error analysis see Appendix B). The results from each method is compared with result from a direct diagonalization of the Hamiltonian, as shown below. From FIG. \ref{result1.png} it can be seen that all of these methods are effective in obtaining the ground state energy problem of the water molecule. We also use method E (Pairwise VQE) to obtain the ground state energy. These results can be seen in FIG. \ref{vqe_result.eps}. Energy convergence at 1.9 a.u. can be seen in FIG. \ref{vqe_v5_iteration.eps} and the ground state energy curve calculated by this method is in FIG. \ref{vqe_energy.eps}. In this simulation,  $d$ is selected to be 1, and $G(\boldsymbol{\theta_i})$ is constructed as described above, and it can already give a very accurate result. This shows Pairwise VQE a very promising method for solving electronic structure problems. Furthermore, Pairwise VQE has only $O(n^2d)$ gate complexity and doesn't require initial input of the ground state, which makes it more practical for near-term applications on a quantum computer.

\begin{figure}[h]
{
\begin{minipage}[t]{0.45\linewidth}
\centering
\includegraphics[width=3in]{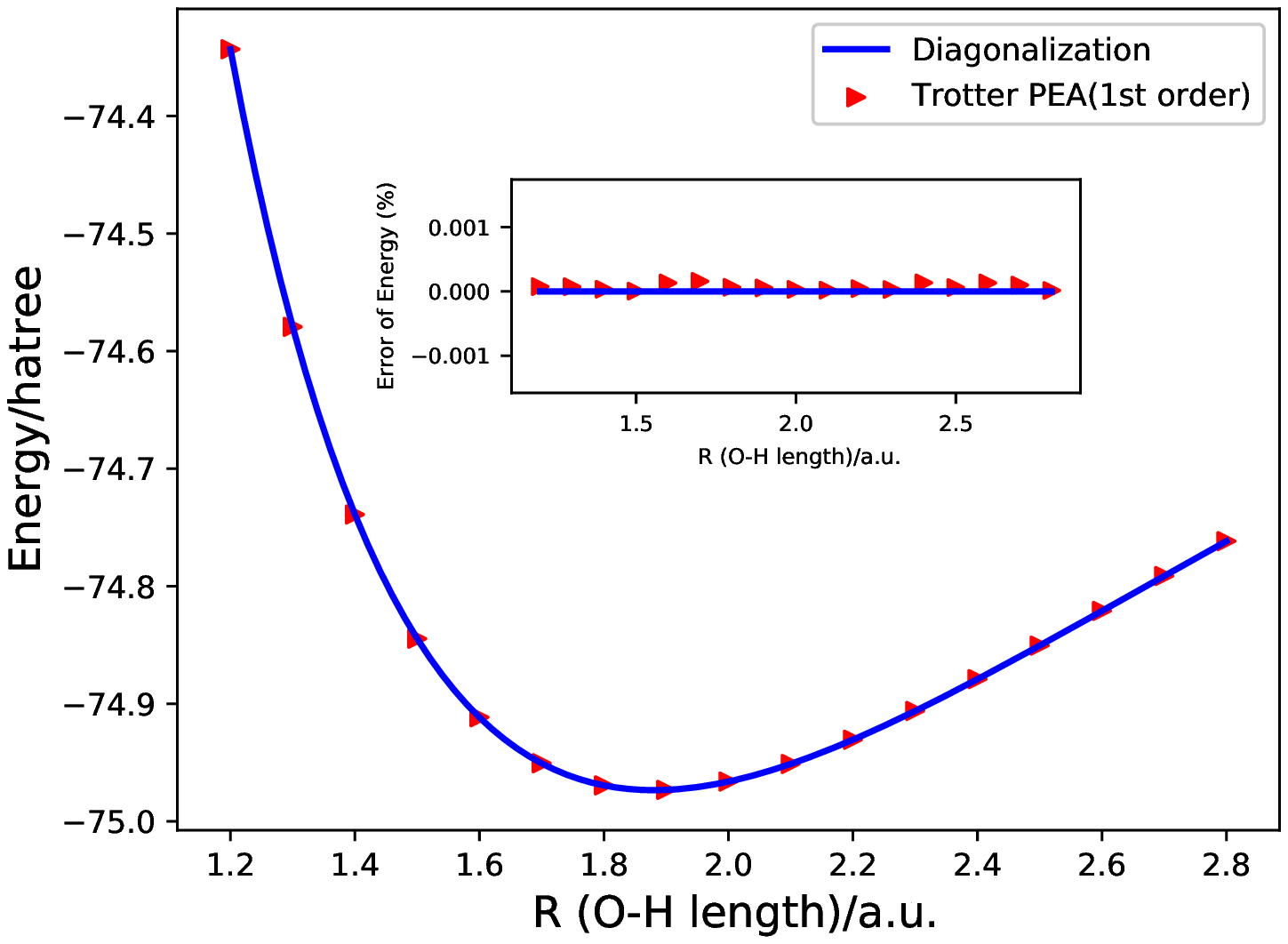}
\caption*{(a)}
\label{fig:side:a}
\end{minipage}%
\begin{minipage}[t]{0.45\linewidth}
\centering
\includegraphics[width=3in]{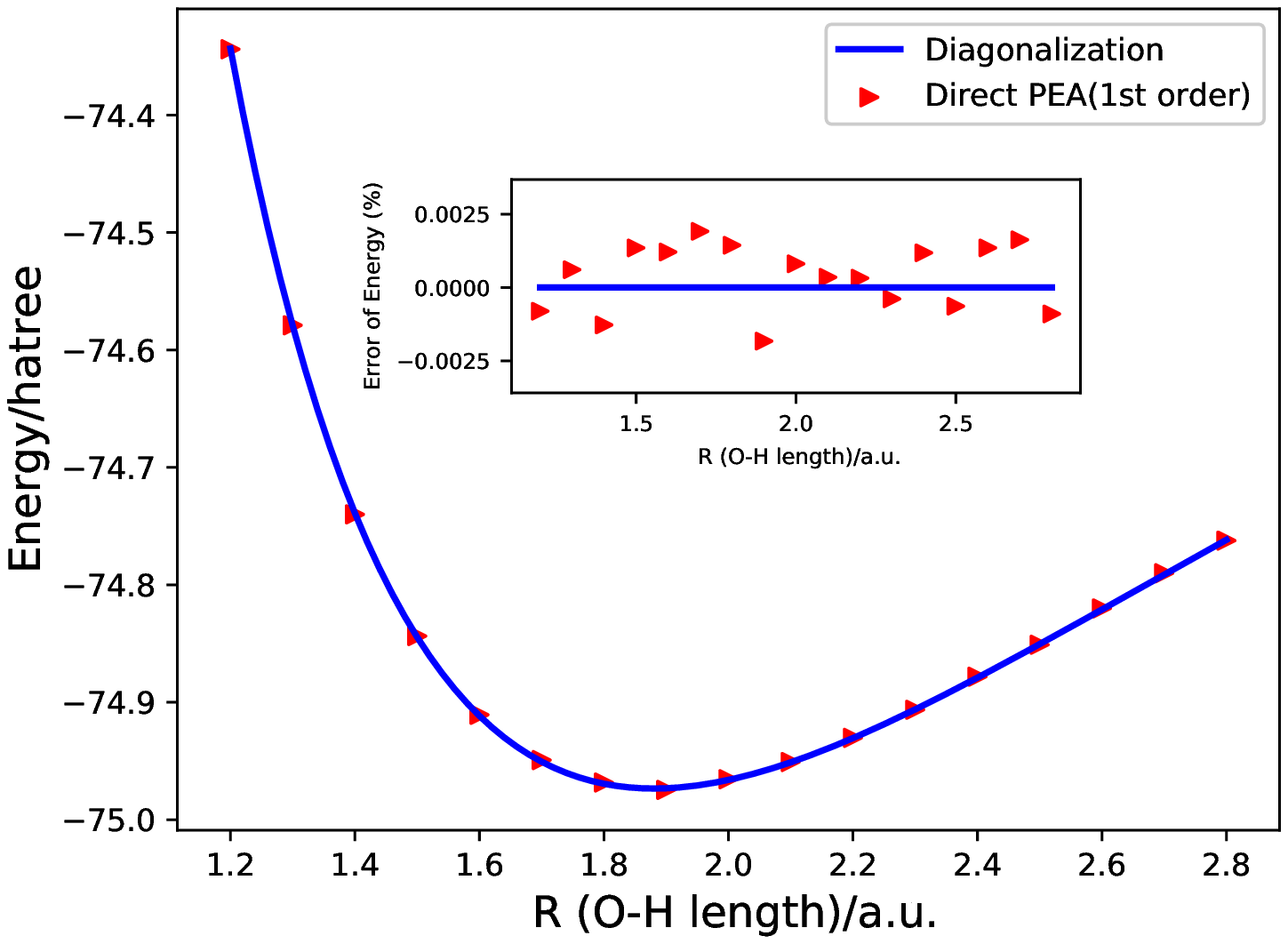}
\caption*{(b)}
\label{fig:side:b}
\end{minipage}
\begin{minipage}[t]{0.45\linewidth}
\centering
\includegraphics[width=3in]{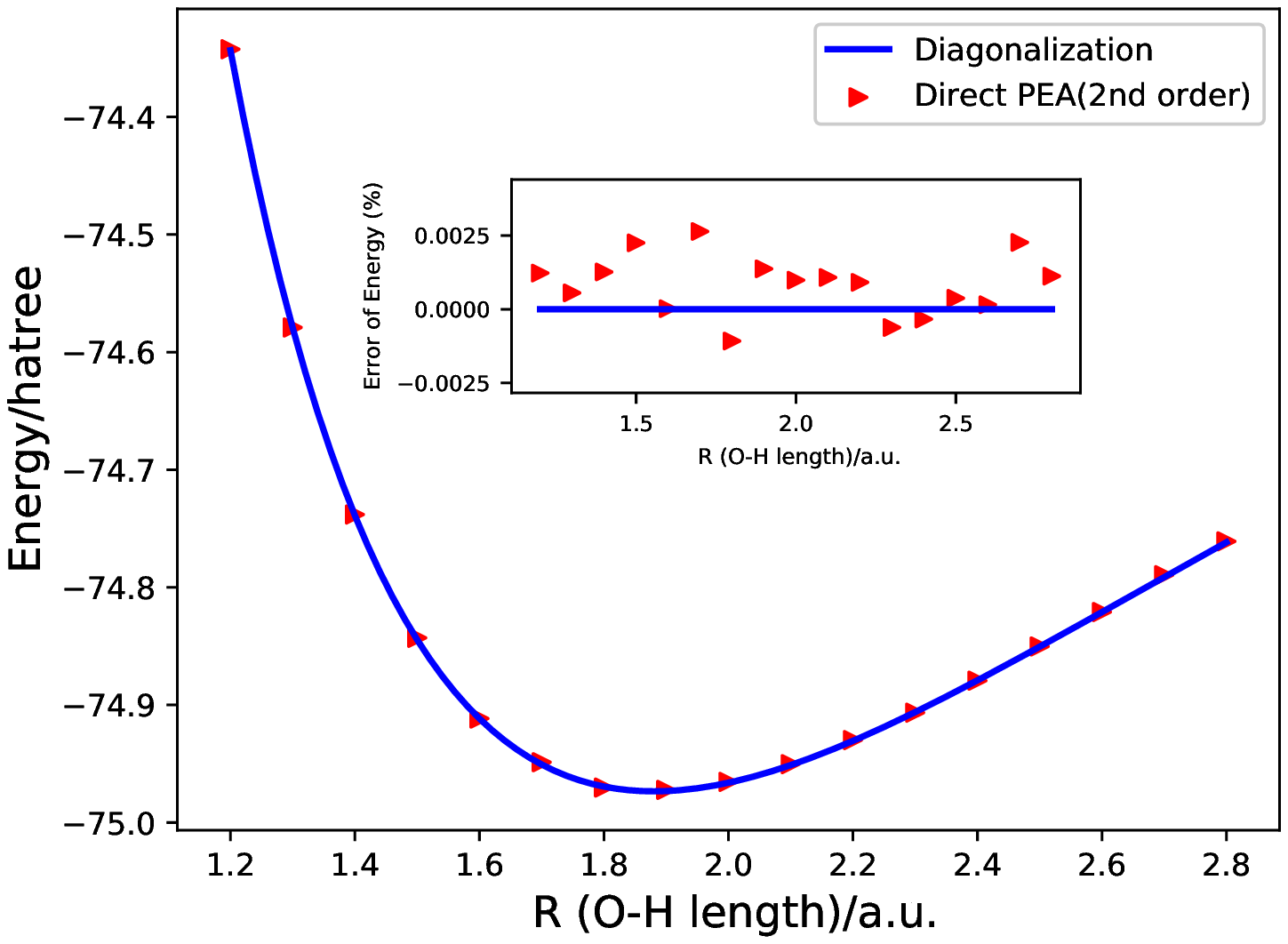}
\caption*{(c)}
\label{fig:side:c}
\end{minipage}
\begin{minipage}[t]{0.45\linewidth}
\centering
\includegraphics[width=3in]{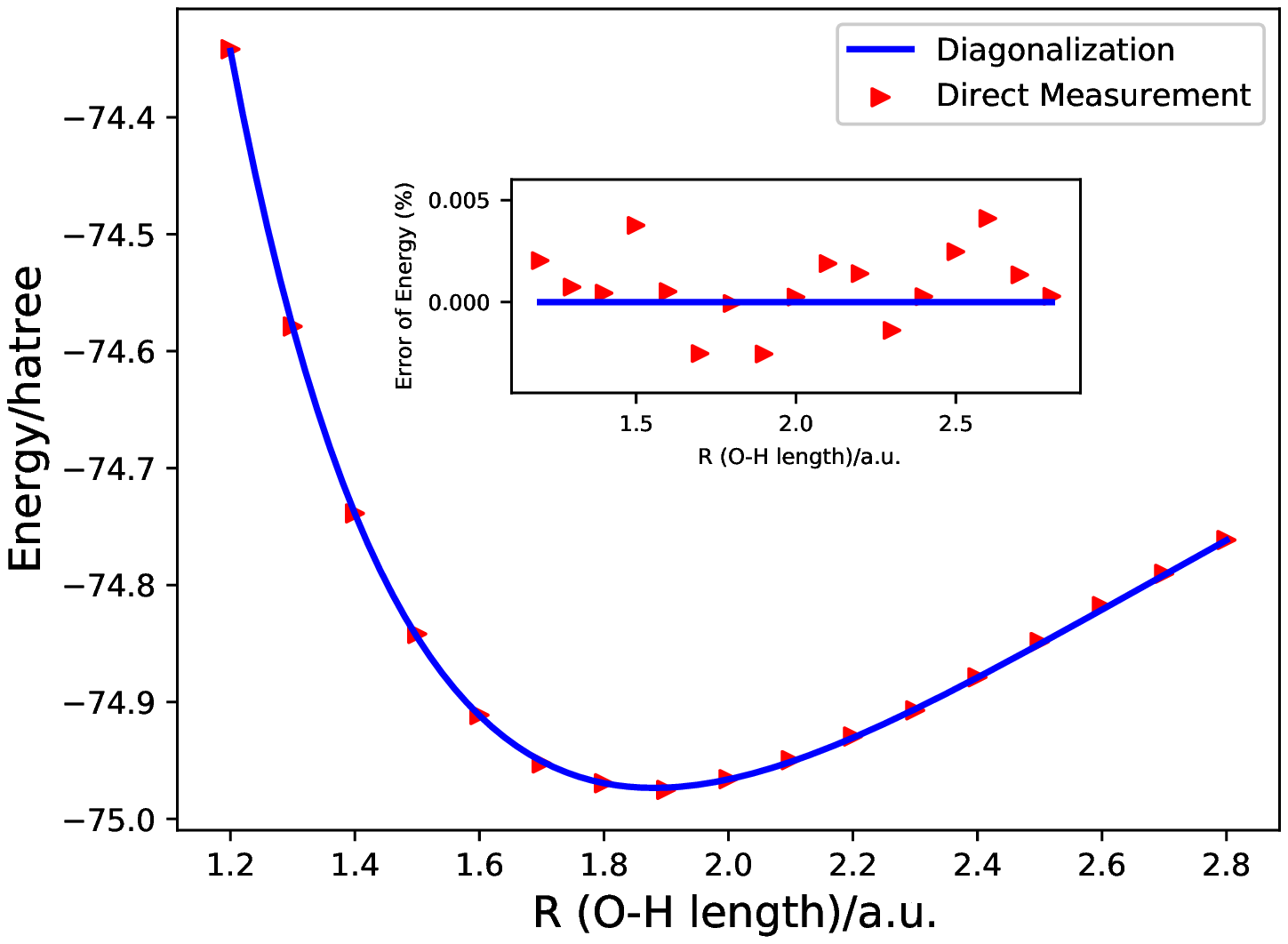}
\caption*{(d)}
\label{fig:side:d}
\end{minipage}
}
\captionsetup{justification=raggedright,singlelinecheck=false}
\caption{
Ground State Energy Curve for H$_2$O, as a function of the bond length O-H in a.u. for (a) the Trotter-PEA, (b) the Direct-PEA ($1^{st}$ order), (c) the Direct-PEA ($2^{nd}$ order) and (d) Direct Measurement method ($1.6\times 10^8$ measurements), compared with the exact diagonalization. Errors are shown in the window of each figure. One thing to mention is that we can not tell whether one method have better property over another directly from these figures, because they have different parameters, gates etc. For comparison, we have to turn to gate complexity analysis in TABLE \ref{complexity_table}.
}

\label{result1.png}
\end{figure}

\begin{figure}
\begin{subfigure}{0.4\linewidth}
    \includegraphics[scale=0.45]{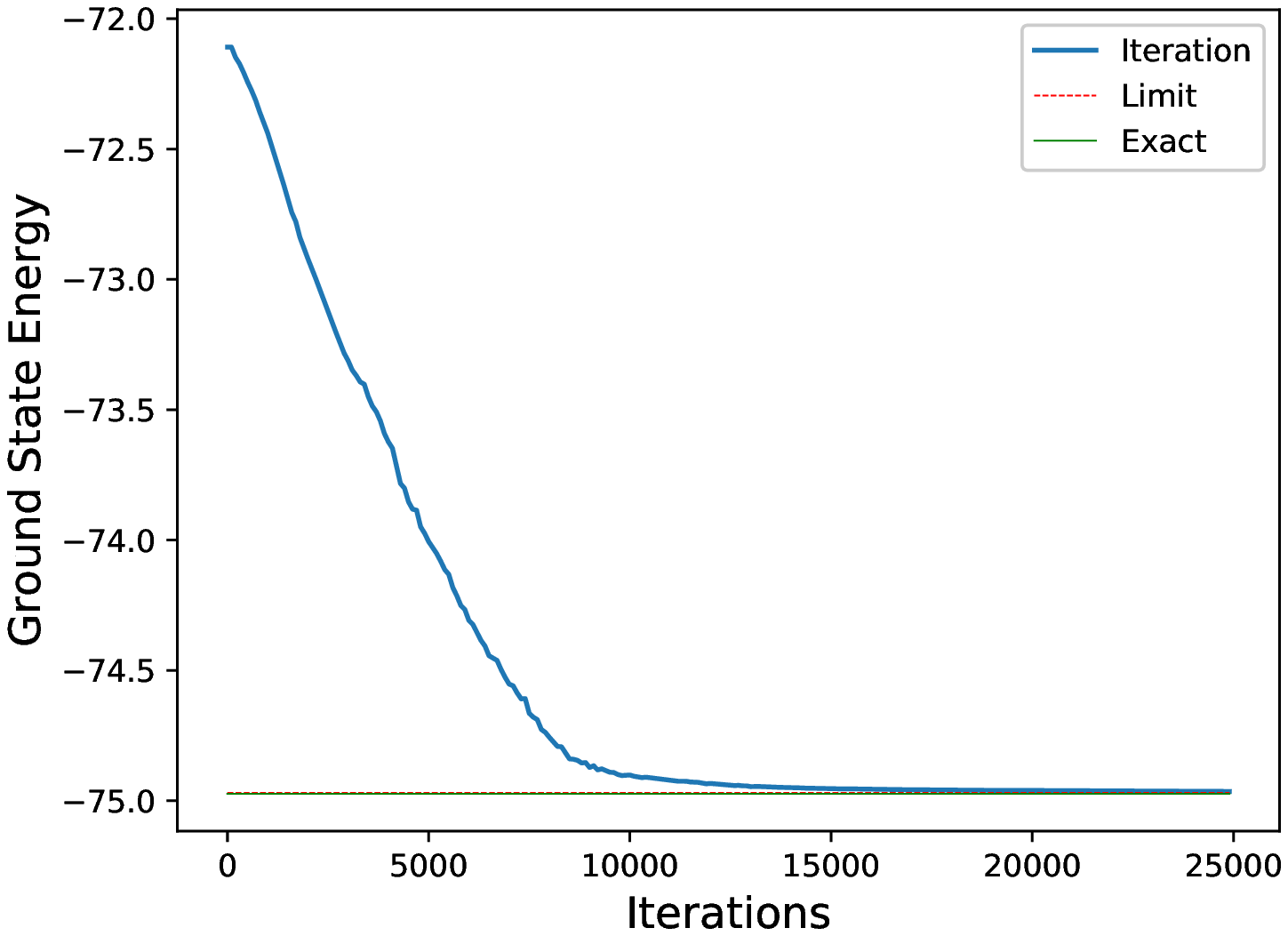}
    \captionsetup{justification=raggedright,singlelinecheck=false}
    \caption{Convergence of ground state energy of H$_2$O for fixed O-H bond length = 1.9 a.u., as number of iterations increases. The lines for exact ground state energy and for the limit almost overlap.}
    \label{vqe_v5_iteration.eps}
\end{subfigure}
\begin{subfigure}{0.4\linewidth}
    \includegraphics[scale=0.45]{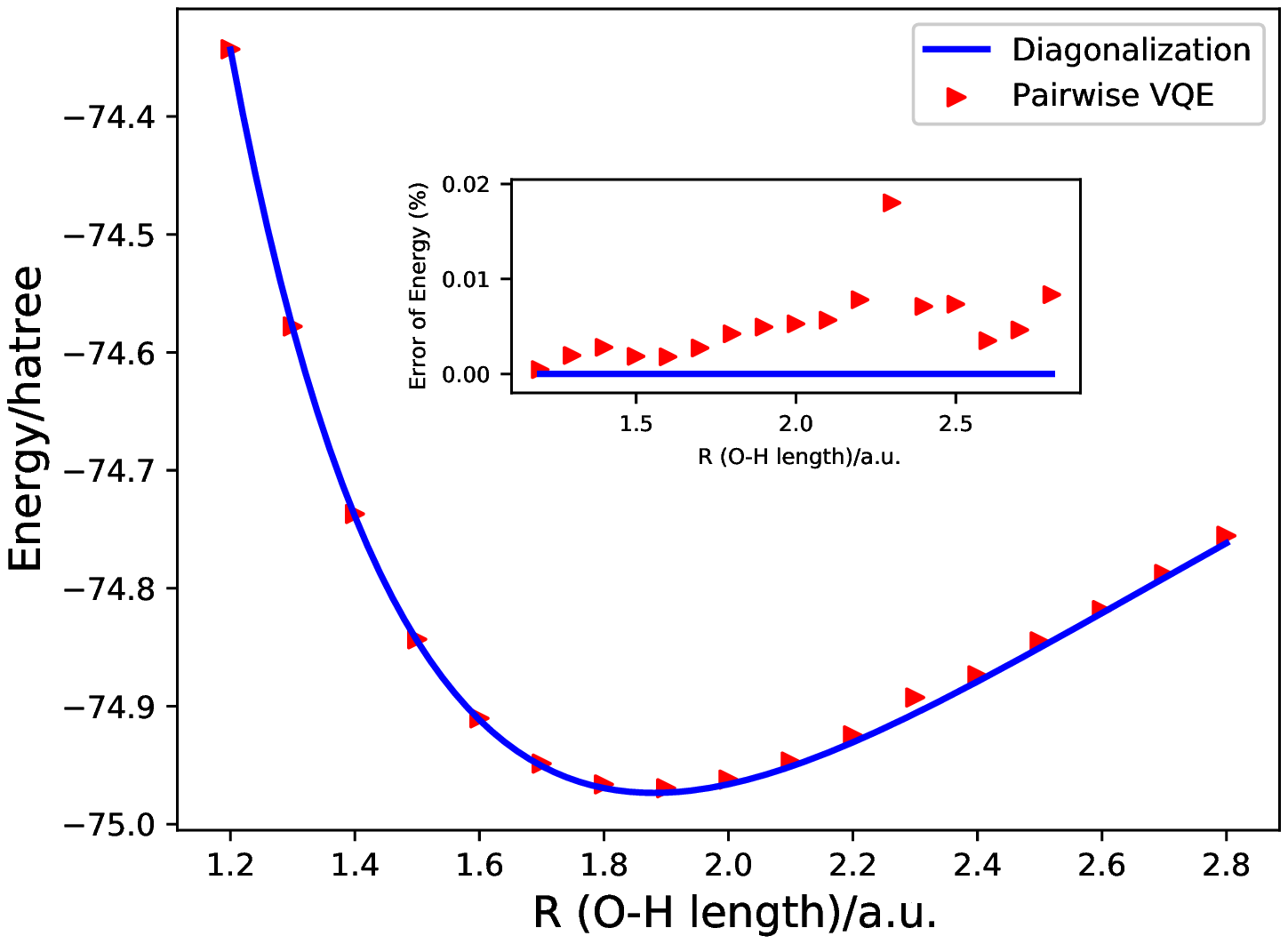}
    \captionsetup{justification=raggedright,singlelinecheck=false}
    \caption{Ground state energy curve for H$_2$O, as a function of O-H bond length in a.u. for variational quantum eigensolver. Errors are shown in the window of the figure.}
    \label{vqe_energy.eps}
\end{subfigure}
\captionsetup{justification=raggedright,singlelinecheck=false}
\caption{Result from Pairwise VQE using the entangling gates in FIG. \ref{vqe_v5.png}. We take $\ket{0}_s$ as initial input, $d=1$ layer and use Nelder-Mead algorithm for optimization. }
\label{vqe_result.eps}
\end{figure}

\newpage
\noindent
Qubit requirement, gate complexity and number of measurements of different methods are analyzed in Appendix C and shown in TABLE \ref{complexity_table}. When counting gate complexities, we decompose all gates into single qubit gates and CNOT gates. While Pairwise VQE needs only n qubits, the other methods require extra number of qubits. In terms of gate scaling, Pairwise VQE also needs the least gates, which enables it to better suit the applications on near and intermediate term quantum computers. Among the remaining four methods, Direct Measurement requires less number of gates than the others. PEA-type methods have an advantage that they can give an accurate result under only $O(1)$ measurements. However, they need more qubits compared with the previous two methods and demands many more gates if smaller error is required. Due to huge gate complexity, these PEA-type algorithms would be put into practice only when the decoherence problem has been better solved. Among these three PEA based methods, in terms of the gate complexity, Direct-PEA(2$^{nd}$ order) requires less number of gates than the traditional Trotter-PEA and Direct-PEA(1$^{st}$ order) which is proved in Appendix C. One more thing to mention is that here the second quantization form Hamiltonian is based on STO-3G, so there are $O(n^4)$ terms. If a more recent dual form of plane wave basis \cite{babbush2017low} is used, the number of terms can be reduced to $O(n^2)$, and the asymptotic scaling in TABLE \ref{complexity_table} would also be reduced. To be specific, for PEA-type methods, upper bounds of gate complexities would be proportional to $n^3$ rather than $n^5$, and Number of Measurements for Pairwise VQE would be proportional to $n^4$ rather than $n^8$. As can be seen, these reductions wouldn't influence the comparison made above.\\

\begin{table}
\begin{tabular}{ |p{3.6cm}||p{3.5cm}|p{3cm}|p{3.9cm}|  }
 \hline
Method & Qubits Requirement & Gate Complexity & Number of Measurements\\
 \hline
Trotter-PEA   & $O(n)$    &$O(\frac{n^5}{(\epsilon/A)^2})$ &   $O(1)$\\
Direct-PEA(1$^{st}$ order)&   $O(n)$  & $O(\frac{n^5}{(\epsilon/A)^{2.5}})$   &$O(1)$\\
Direct-PEA(2$^{nd}$ order)&   $O(n)$  & $O(\frac{n^5}{(\epsilon/A)^{1.3}})$   &$O(1)$\\
Direct Measurement & $O(n)$ & $O(n^5)$ & $O(\frac{E^2}{\epsilon^2}) $\\
Pairwise VQE & $n$ & $O(n^2d)$ & $O(\frac{A^2n^8}{\epsilon^2}N_{iter})$\\
 \hline
\end{tabular}
\captionsetup{justification=raggedright,singlelinecheck=false}
\caption{Complexity of different methods. $n$ is the number of qubits for molecular system, 6 for water in this paper. $ A = \sum_{i=1}^L |\alpha_i|$ can serve as the scale of energy. E is the exact value of ground energy. $\epsilon$ is the accuracy of energy we want to reach. $d$ is the number of layers we used in Pairwise VQE. $N_{iter}$ is the number of iterations for optimization in Pairwise VQE. See Appendix C for details.}
\label{complexity_table}
\end{table}

\section{Excited states and resonances}

All the aforementioned methods can also be applied for the excited state energy calculation. For PEA-type methods and Direct Measurement method, it can be simply done by replacing the input system state by an excited state. The complexity for the calculation is the same. The energy accuracies for excited states are also similar to that for the ground state. For VQE, a recent publication \cite{colless2018computation,sim2018quantum} presents a  quantum subspace expansion algorithm (QSE) to calculate excited state energies. They approximate a ``subspace'' of low-energy excited states from linear combinations of states of the form $O_i\ket{\psi}_s$, where $\ket{\psi}_s$ is the ground state determined by VQE and $O_i$ are chosen physically motivated quantum operators. By diagonalizing the matrix with elements $\bra{\psi}_sO_i^\dagger H O_j \ket{\psi}_s$ calculated by VQE, one is able to find the energies of excited states.

\begin{figure}[]
{
\includegraphics[]{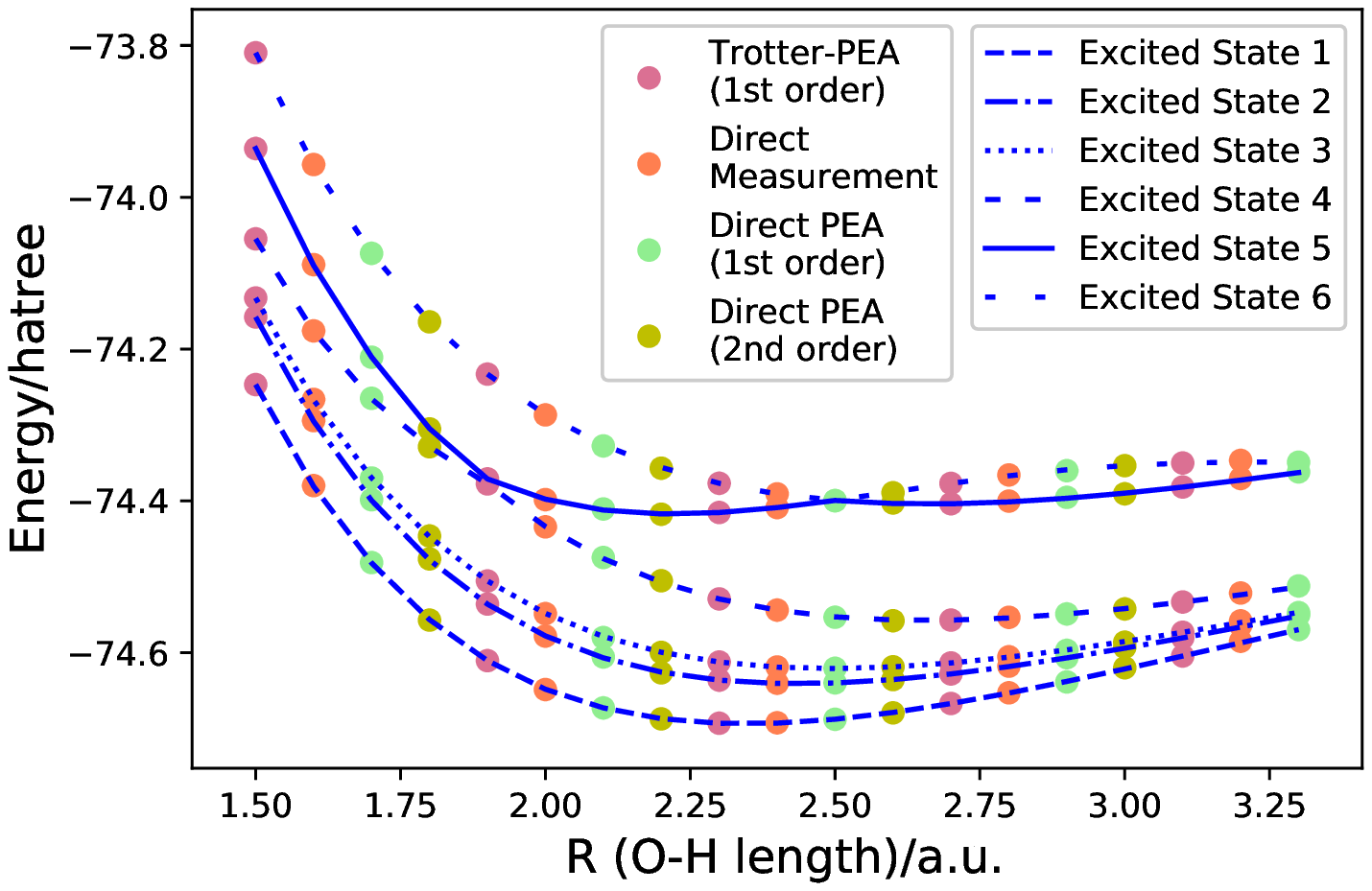}
}
\captionsetup{justification=raggedright,singlelinecheck=false}
\caption{
Excited states' energy curves for H$_2$O, as a function of the bond length O-H in a.u.. Markers with different colors represent data points calculated from different methods. Only a few points for each method are drawn for illustration. Energy curves in different line styles are calculated from exact diagonalization of Hamiltonian matrix.}
\label{result_excited.png}
\end{figure}

FIG. \ref{result_excited.png} shows the simulation of the first six excited states' energy curves of the water molecule from our 6-qubit Hamiltonian, calculated by PEA-type methods and Direct Measurement method. It can be seen that the $5^{th}$ excited energy curve indicates a shape resonance phenomenon, which can be described by a non-Hermitian Hamiltonian with complex eigenvalues. The life time of the resonance state is associated with the imaginary part of the eigenvalues. In this way, to solve the resonance problem, we can seek to solve the eigenvalues of non-Hermitian Hamiltonians.

Some work has been done on this track to solve the resonance problem by quantum computers. By designing a general quantum circuit for non-unitary matrices, Daskin et al.\cite{daskin2014universal} explored the resonance states of a model non-Hermitian Hamiltonian. To be specific, he introduced a systematic way to estimate the complex eigenvalues of a general matrix using the standard iterative phase estimation algorithm with a programmable circuit design. The bit values of the phase qubit determines the phase of eigenvalue, and the statistics of outcomes of the measurements on the phase qubit determines the absolute value of the eigenvalue. Other approaches for solving complex eigenvalues can also be applied for this resonance problem. For example, Wang et al. \cite{wang2010measurement} proposed a measurement-based quantum algorithms for finding eigenvalues of non-unitary matrices. Terashima et al.\cite{terashima2005nonunitary} introduced a universal nonunitary quantum circuit by using a specific type of one-qubit non-unitary gates, the controlled-NOT gate, and all one-qubit unitary gates, which is also useful for finding the eigenvalues of a non-hermitian Hamiltonian matrix.

Method D in section III can also be used for solving complex eigenvalues and the complexity is polynomial in system size. After applying complex-scaling method\cite{simon1973resonances} to water molecule's Hamiltonian and obtaining a non-Hermitian Hamiltonian, we can make enough quantum measurements to get an accurate resonance width $\Gamma$, which is actually the imaginary part of Hamiltonian's eigenvalue\cite{moiseyev1979autoionizing}. Another easier way to solve this resonance problem is, we can first choose proper $a$ and $J$ to fit the potential energy in a widely studied Hamiltonian \cite{moiseyev1978resonance,moiseyev1984resonances,serra2001crossover}:
 \begin{align}
     H(x) = \frac{p^2}{2} + (\frac{x^2}{2}-J)e^{-ax^2}
 \end{align}
 to our energy curve. Then by complex-scaling method, the internal coordinates of the Hamiltonian is dilated by a complex factor $\eta = \alpha e^{-i\theta}$ such that $H(x) \rightarrow H(x/\eta) \equiv H_\eta(x)$. We can solve the complex eigenvalue of $H_{\eta}(x)$ by the method D or using our previous method \cite{daskin2014universal}.

\section{Conclusion}

In this study we have compared several recently proposed quantum algorithms when used to compute the electronic state energies of the water molecule. These methods include first order Trotter-PEA method based on the first order Trotter decomposition, first and second order Direct-PEA methods based on direct implementation of the truncated propagator, Direct Measurement method based on direct implementation of the Hamiltonian and Pairwise PEA method, a VQE algorithm with a designed ansatz. 

After deriving the Hamiltonian of the water molecular using the STO-3G basis set, we have explained in detail how each method works and derived their qubit requirements, gate complexities and measurement scalings. We have also calculated the ground state energy of the water molecular and shown the ground energy curves from all five methods. All methods are able to provide an accurate result. We have compared these methods and concluded that the second order Direct-PEA provides the most efficient circuit implementations in terms of gate complexity. With large scale quantum computation, the second order direct method seems to better suit large molecule systems. In addition,  since Pairwise VQE requires the least qubit number, it is the most practical method for near-term applications on the current available quantum computers.

Moreover, we have applied our PEA-type methods and Direct Measurement method to solve excited state energy curves for water molecule.  The fifth excited state energy curve implies shape resonance. We have introduced recent work on quantum algorithms for solving the molecular resonance problems and given two possible ways to solve the water molecule resonance properties, including our Direct Measurement method which is able to solve the problem efficiently.\\

\noindent
\textbf{Acknowledgements}:\\
This material is based upon work supported by the U.S. Department of Energy, Office of Basic Energy Sciences, under Award Number DE-SC0019215. Author Daniel Murphy acknowledges the support of NSF REU award PHY-1460899.

\bibliography{ref}

\newpage
\section*{Appendix A: H$_2$O Hamiltonian at Equilibrium}
\begin{center}
\begin{tabular}{||c | c||c | c|| c | c||}
\hline
IIIIII & -72.008089 & IIIIIZ & 0.373979 & IIIIXX & -0.050755 \\ 
IIIIYY & 0.113535 & IIIIZI & 0.002526 & IIIIZZ & 0.779273 \\ 
IIIZII & -0.771553 & IIIZIZ & 0.043092 & IIIZXX & 0.113535 \\ 
IIIZYY & -0.050755 & IIIZZI & 0.785287 & IIIZZZ & -0.030367 \\ 
IIXIIX & 0.009295 & IIXIXI & 0.000158 & IIXIZX & -0.009295 \\ 
IIXZXZ & -0.000158 & IIZIII & -0.373979 & IIZIIZ & -0.148141 \\ 
IIZIYY & -0.011744 & IIZIZZ & -0.146285 & IIZZII & 0.141059 \\ 
IIZZXX & -0.011744 & IIZZZI & -0.136887 & IXIIIX & 0.000158 \\ 
IXIIXI & 0.013400 & IXIIZX & -0.000158 & IXIZXZ & -0.013400 \\ 
IXXIII & -0.050755 & IYYIII & 0.113535 & IYYIIZ & 0.011744 \\ 
IYYIYY & 0.019371 & IYYIZZ & 0.031747 & IYYZII & -0.011216 \\ 
IYYZXX & 0.019371 & IYYZZI & 0.031561 & IZIIII & -0.002526 \\ 
IZXIIX & 0.009295 & IZXIXI & 0.000158 & IZXIZX & -0.009295 \\ 
IZXZXZ & -0.000158 & IZZIII & 0.779273 & IZZIIZ & 0.146285 \\ 
IZZIYY & 0.031747 & IZZIZZ & 0.220040 & IZZZII & -0.154863 \\ 
IZZZXX & 0.031747 & IZZZZI & 0.179396 & XIIXII & 0.012412 \\ 
XIIXXX & -0.007950 & XIIXZI & 0.012412 & XIIYXY & 0.007950 \\ 
XXXXII & -0.007950 & XXXXXX & 0.018156 & XXXXZI & -0.007950 \\ 
XXXYXY & -0.018156 & XXZXXZ & -0.006979 & XXZYYI & 0.006979 \\ 
XZIXII & -0.012412 & XZIXXX & 0.007950 & XZIXZI & -0.012412 \\ 
XZIYXY & -0.007950 & YXYXII & 0.007950 & YXYXXX & -0.018156 \\ 
YXYXZI & 0.007950 & YXYYXY & 0.018156 & YYIXXZ & -0.006979 \\ 
YYIYYI & 0.006979 & ZIIIII & 0.771553 & ZIIIIZ & 0.141059 \\ 
ZIIIYY & 0.011216 & ZIIIZZ & 0.154863 & ZIIZII & -0.154860 \\ 
ZIIZXX & 0.011216 & ZIIZZI & 0.146877 & ZIZIII & 0.043092 \\ 
ZXXIII & -0.113535 & ZXXIIZ & -0.011744 & ZXXIYY & -0.019371 \\ 
ZXXIZZ & -0.031747 & ZXXZII & 0.011216 & ZXXZXX & -0.019371 \\ 
ZXXZZI & -0.031561 & ZXZIIX & -0.000158 & ZXZIXI & -0.013400 \\ 
ZXZIZX & 0.000158 & ZXZZXZ & 0.013400 & ZYYIII & 0.050755 \\ 
ZZIIII & 0.785287 & ZZIIIZ & 0.136887 & ZZIIYY & 0.031561 \\ 
ZZIIZZ & 0.179396 & ZZIZII & -0.146877 & ZZIZXX & 0.031561 \\ 
ZZIZZI & 0.189343 & ZZZIII & 0.030367 & & \\
\hline
\end{tabular}
\captionof{table}{Spin-type Hamiltonian of the water molecule at equilibrium when O-H is 1.9 a.u. There are 95 terms, and listed are each operator and corresponding coefficient. $X,Y,Z,I$ stand for the spin matrices $\sigma^x,\sigma^y,\sigma^z$ and the identity operator on a single qubit subspace.}
\end{center}

\newpage
\section*{Appendix B. Error Analysis}

\subsection*{B.1 Trotter PEA}
\noindent
The Trotter decomposition is 
\begin{align}
    e^{-iHt} = \prod\limits_{i=1}^L e^{-i\alpha_i h_i t} + O(A^2t^2)
\end{align}

\noindent
Suppose our input is an eigenstate of $\ket{\varphi}_s$ of the Hamiltonian and has $H\ket{\varphi}_s = E\ket{\varphi}_s$, then:
\begin{align}
    \prod\limits_{i=1}^L e^{-i\alpha_i h_i t} \ket{\varphi}_s &= (e^{-iHt} - O(A^2t^2) \ket{\varphi}_s \nonumber\\
    & = (e^{-iEt} - O(A^2t^2)) \ket{\varphi}_s + O(A^2t^2) \ket{\varphi^\perp} \nonumber\\
    &= e^{-iEt} (1 - O(A^2t^2)e^{iE}) \ket{\varphi}_s + O(A^2t^2) \ket{\varphi^\perp} \nonumber\\
    & = e^{-iEt} (1 - O(A^2t^2) - iO(A^2t^2)) \ket{\varphi}_s + O(A^2t^2) \ket{\varphi^\perp} \nonumber\\
    & = (1-O(A^2t^2))e^{-iEt} e^{i \tan^{-1}(\frac{O(A^2t^2)}{1-O(A^2t^2)})}\ket{\varphi}_s + O(A^2t^2) \ket{\varphi^\perp} \nonumber\\
    & = (1-O(A^2t^2))e^{-i(Et+O(A^2t^2)}\ket{\varphi}_s + O(A^2t^2) \ket{\varphi^\perp}
\end{align}
\noindent
It should be noticed that in this equation, $O(A^2t^2)$ is an operator before being applied to $\ket{\varphi}_s$.\\

\noindent
In this way, the possibility that we can measure the correct ground state energy is $1-O(A^2t^2)$. After $2^D$ gates, in which $D$ represents the number of digits we want to measure by PEA, the probability of state $\ket{0}\ket{\psi}_s$ should be still large. By setting the final coefficient to be $1-\frac{1}{8}$, then:
\begin{align}
    (1-O(A^2t^2))^{2^D} = 1-\frac{1}{8}\\
    2^{-D} = O(A^2t^2)
\end{align}

\noindent
The error of the energy resulting from the phase is: $\epsilon_1 = O(A^2t^2)$. If we use PEA until $D$ digits, the error of the energy resulting from PEA is: $\epsilon_2 = O(2^{-D}/t) = O(A^2t)$. Then totally we have an error: $\epsilon = O(\epsilon_1+\epsilon_2) = O(A^2t)$.

\noindent
Since the error for the first order Trotter-Suzuki decomposition is: 
\begin{align}
    e^{-iHt} - \prod_{i=0}^Le^{-\frac{i\alpha_ih_it}{2}} \prod_{i'=L}^0 e^{-\frac{i\alpha_i' h_{i'} t}{2}} = O(A^3t^3),    
\end{align}
\noindent
by a similar analysis the total error after PEA based on Trotter-Suzuki decomposition would be $\epsilon = O(A^3t^2)$.

\subsection*{B.2 Direct PEA (First Order)}

\noindent
From the main part, after gate $U_r$, we obtain:
\begin{align}
    U_r \ket{0}_a \ket{\psi}_s &= \frac{\sqrt{1+\frac{E^2}{\kappa^2}}}{s} e^{-i\tan^{-1} \frac{E}{\kappa}} \ket{0}_a \ket{\psi}_s + \ket{\Phi^\perp_1} \nonumber\\
    &= \frac{\sqrt{1+\frac{E^2}{\kappa^2}}}{1+\frac{A}{\kappa}} e^{-i\tan^{-1} \frac{E}{\kappa}} \ket{0}_a \ket{\psi}_s + \ket{\Phi^\perp_1} \nonumber\\
    &= pe^{-i\tan^{-1}\frac{E}{\kappa}} \ket{0}_a \ket{\psi}_s + \sqrt{1-p^2} \ket{\Phi^\perp}  \nonumber\\
    &= \cos\theta e^{-i\tan^{-1} \frac{E}{\kappa}} \ket{0}_a \ket{\psi}_s + \sin\theta  \ket{\Phi^\perp}
\end{align}
Here $A = \sum_{i=1}^{2^m-1} \beta_i= \sum_{i=1}^L |\alpha_i| \geq |E|$, $\theta = \arccos \frac{\sqrt{1+\frac{E^2}{\kappa^2}}}{1+\frac{A}{\kappa}}$. \\

\noindent
To increase the probability of $\ket{0}_a \ket{\psi}_s$, we use $Q = U_r (U_0 \otimes I^{\otimes n}) U_r^\dagger (U_0 \otimes I^{\otimes n})$ to do oblivious amplitude amplification:
\begin{align}
    Q^N U_r\ket{0}_a \ket{\psi}_s &= (-1)^N\cos((2N+1)\theta) e^{-i\tan^{-1} \frac{E}{\kappa}} \ket{0}_a \ket{\psi}_s + \sin((2N+1)\theta)  \ket{\Phi^\perp} \nonumber\\
    &=p_f \ket{0}_a \ket{\psi}_s + \sqrt{1-p_f^2}  \ket{\Phi^\perp}
\end{align}

\noindent
The idea is, if $\kappa$ is large, $\frac{\sqrt{1+\frac{E^2}{\kappa^2}}}{1+\frac{A}{\kappa}} \approx \frac{1}{1+\frac{A}{\kappa}}$, and $\theta' =  \cos^{-1} \frac{1}{1+\frac{A}{\kappa}} \approx \theta$. By choosing large $N$ and $\kappa$ to satisfy $(2N+1)\theta' = \pi$, which means $\frac{A}{\kappa} = \frac{1}{\cos(\frac{\pi}{2N+1})}-1$, we are able to get $\cos((2N+1)\theta) \approx -1$.\\

\noindent
Since:
\begin{align}
    \theta - \theta' &= \cos^{-1}(\frac{\sqrt{1+\frac{E^2}{\kappa^2}}}{1+\frac{A}{\kappa}}) - \cos^{-1}(\frac{1}{1+\frac{A}{\kappa}}) \nonumber\\
    &= \frac{\sqrt{2}}{4}\eta^2 (\frac{A}{\kappa})^{\frac{3}{2}} + O((\frac{A}{\kappa})^{\frac{5}{2}})
\end{align}
In which $\eta = |\frac{E}{A}| \leq 1 $. Then after $N$ rotations
\begin{align}
    |p_f| &= |\cos((2N+1)\theta)| \nonumber\\
    &= \cos((2N+1)(\theta'-\theta)) \nonumber\\
    &= 1-\frac{(2N+1)^2}{16}\eta^4(\frac{A}{\kappa})^3 + O((\frac{A}{\kappa})^4) \nonumber\\
    &= 1 - \frac{\pi^6}{2^{11}}\eta^4\frac{1}{N^4} + O(\frac{1}{N^5})
\end{align}

\noindent
This means if we set large enough $N$, and then set $\kappa = \frac{A\cos(\frac{\pi}{2N+1})}{1-\cos(\frac{\pi}{2N+1})}$, we are able to amplify the probability of $\ket{0}_a\ket{\psi}_s$ to be as close to 1 as we want.\\

\noindent
Now we are taking $U_q = Q^NU_r$ to encode the energy into the phase. For the next PEA step, if we would like $D$ digit accuracy, we have to make sure after $2^D$ gates of $U_q$, the probability of state $\ket{0}_a\ket{\psi}_s$ is still large. To make the analysis easier, we set the final coefficient for that state  $1-\frac{1}{2^3}$. Then the following formula should be satisfied:
\begin{align}
    &|p_f|^{2^D} = 1-\frac{1}{2^3}\\
    &2^{-D} = \frac{\pi^6\eta^4}{2^{11}\ln(\frac{8}{7})}\frac{1}{N^4} + O(\frac{1}{N^5})\\
    &D = \min\{ \log_2(\frac{2^{11}\ln(\frac{8}{7})}{\pi^6\eta^4})+4\log_2 N\} \approx -1.81+4\log_2 N
\end{align}
Since D-digit output from PEA gives us the phase $\varphi$ to approximate $\frac{1}{2\pi}\tan^{-1}\frac{-E}{\kappa}$, and the error of the phase is $2^{-D}$, we get the error of the energy to be:
\begin{align}
    \epsilon &= \tan(2\pi*2^{-D})\times \kappa = \frac{\pi^5\eta^4}{2^7\ln(\frac{8}{7})}\frac{1}{N^2}+O(\frac{1}{N^3}) \nonumber\\
    &\approx \frac{17.90\eta^4A}{N^2} \leq \frac{17.90}{N^2}A
\end{align}

\noindent
We can see that, by taking large N and set corresponding $\kappa$ (which is also large), we are able to control the accuracy of PEA process.

\subsection*{B.3 Direct PEA (Second Order)}
\noindent
From the main part, after gate $U_{2r}$, we obtain:
\begin{align}
    U_{r2}\ket{00}\ket{0}_a\ket{\psi}_s &= \frac{\sqrt{1+\frac{E^4t^4}{4A^2}}}{1+t+\frac{t^2}{2}} e^{-i\tan^{-1}{\frac{\frac{Et}{A}}{1+\frac{E^2t^2}{2A}}}} \ket{00}\ket{0}_a\ket{\psi}_s + \ket{\Psi_1^\perp} \nonumber\\
    &= p e^{-i\tan^{-1}{\frac{\frac{Et}{A}}{1+\frac{E^2t^2}{2A}}}} \ket{00}\ket{0}_a\ket{\psi}_s + \sqrt{1-p^2}\ket{\Psi^\perp} \nonumber\\
    &= \cos \theta e^{-i\tan^{-1}{\frac{\frac{Et}{A}}{1+\frac{E^2t^2}{2A}}}} \ket{00}\ket{0}_a\ket{\psi}_s + \sin \theta \ket{\Psi^\perp}
\end{align}
Here $A = \sum_{i=1}^{2^m-1}\beta_i = \sum_{i=1}^L |\alpha_i| \geq |E|, \theta = \cos^{-1}{\frac{\sqrt{1+\frac{E^4t^4}{4A^2}}}{1+t+\frac{t^2}{2}}}$.

\noindent
Apply $Q_2 = U_{2r}(U_0^+\otimes I^{\otimes n})U_{2r}^{\dagger}(U_0^+\otimes I^{\otimes n})$, in which $U_0^+ = 2\ket{00}\ket{0}_a\bra{0}_a\bra{00}-I^{\otimes m+2}$, to do obivious amplitude amplification:
\begin{align}
    Q_2^N U_{2r}\ket{00}\ket{0}_a \ket{\psi}_s &= (-1)^N\cos((2N+1)\theta) e^{-i\tan^{-1}{\frac{\frac{Et}{A}}{1+\frac{E^2t^2}{2A}}}} \ket{0}_a \ket{\psi}_s + \sin((2N+1)\theta)  \ket{\Psi^\perp}_{a+s+2} \nonumber\\
    &=p_f \ket{00}\ket{0}_a \ket{\psi}_s + \sqrt{1-p_f^2}  \ket{\Psi^\perp}_{a+s+2}
\end{align}

\noindent
Let $\theta' = \cos^{-1}\frac{1}{1+t+\frac{t^2}{2}}$ and choose large $N$ and small $t$ to satisfy $(2N+1)\theta' = \pi$, which leads to 
\begin{align}
    t &= -1+\sqrt{\frac{2}{\cos{\frac{\pi}{2N+1}}}-1} = \frac{\pi^2}{8N^2} + O(\frac{1}{N^3})
\end{align}

\noindent
Then:
\begin{align}
    \theta-\theta' &=\cos^{-1}{\frac{\sqrt{1+\frac{E^4t^4}{4A^2}}}{1+t+\frac{t^2}{2}}} - \cos^{-1}{\frac{1}{1+t+\frac{t^2}{2}}} = \frac{\sqrt{2}}{16}\eta^4 t^{\frac{7}{2}} + O(t^{\frac{9}{2}})
\end{align}

\noindent
In wihch $\eta = |\frac{E}{A}| \leq 1$. Then after $N$ rotations
\begin{align}
    |p_f| &= |\cos((2N+1)\theta)| \nonumber\\
    &= \cos((2N+1)(\theta'-\theta)) \nonumber\\
    &= 1-\frac{(2N+1)^2}{16^2}\eta^8t^7 + O(t^8) \nonumber\\
    &= 1- \frac{\pi^{14}}{2^{27}}\eta^8\frac{1}{N^{12}} + O(\frac{1}{N^{13}})
\end{align}

\noindent
This means if we set large enough N, and then set $t = -1+\sqrt{\frac{2}{\cos{\frac{\pi}{2N+1}}}-1}$, we are able to amplify the probability of $\ket{00}\ket{0}_a\ket{\psi}_s$ to be as close to 1 as we want.

\noindent
Now we are taking $U_{q2} = Q_2^N U_{r2}$ to encode the energy into the phase, if we would like $D$ digit accuracy, we have to make sure after $2^D$ gates of $U_{q2}$, the probability of state $\ket{0}_a\ket{\psi}_s$ is still large. By setting the final coefficient is $1-\frac{1}{2^3}$, then the following formula should be satisfied:

\begin{align}
    &|p_f|^{2^D} = 1-\frac{1}{2^3}\\
    &2^{-D} = \frac{\pi^{14}\eta^8}{2^{27}\ln(\frac{8}{7})}\frac{1}{N^{12}} + O(\frac{1}{N^{13}})\\
    &D = \min\{ \log_2(\frac{2^{27}\ln(\frac{8}{7})}{\pi^{14}\eta^8})+12\log_2 N\} \approx 0.974+12\log_2 N
\end{align}

\noindent
Since D-digit output from PEA gives us the phase $\varphi$ to approximate $-\frac{1}{2\pi} \tan^{-1}{\frac{\frac{Et}{A}}{1+\frac{E^2t^2}{2A}}} $ and the error of phase is $2^{-D}$, we get the error of the energy $E$ to be:
\begin{align}
    \epsilon = \frac{\pi^{13}\eta^8}{2^{23}\ln{\frac{8}{7}}} \frac{A}{N^{10}} + O(\frac{A}{N^{11}}) \approx \frac{2.59\eta^8A}{N^{10}} \leq \frac{2.59}{N^{10}}A
\end{align}

\noindent
We can see that by taking large $N$ and set corresponding small $t$, we are able to control the accuracy of PEA process.

\subsection*{B.4 Direct Measurement}
\noindent
After applying the gate $U_r'$:
\begin{align}
    U_r'\ket{0}_a\ket{\psi}_s = \frac{E}{A}\ket{0}_0\ket{\psi}_s + \ket{\Phi_1^{'\perp}}
\end{align}

\noindent
We obtain eigenenergy of state $\ket{\psi}_s$ by calculating probability of the wanted state: $\ket{0}_a\ket{\psi}_s$. The standard error of $E$ by $X$ measurements is:
\begin{align}
    \sigma = \frac{|E|}{\sqrt{X}} \sqrt{1-\frac{E^2}{A^2}}
\end{align}

\section*{Appendix C. Complexity}
\subsection*{C.1 Trotter PEA}
\noindent
We need $n$ qubits for the state and at least 1 qubit for PEA process, so totally we need $O(n)$ qubits.\\

\noindent
If we measure the ground state energy to $D$ bit precision, we need $O(2^DL n)$ standard gates to implement PEA, in which by saying standard gates we mean single qubit gates and CNOT gates. Since $2^D = O(\frac{1}{A^2t^2}) = O(\frac{A^2}{\epsilon^2})$, and for molecular system, $L=n^4$, so the gate complexity of Trotter PEA would be $O(\frac{n^5}{(\epsilon/A)^2})$.\\

\noindent
For PEA based on the second order Trotter-Suzuki decomposition, we still need $O(2^DLn)$ standard gates. Now $2^D = O(\frac{1}{A^3t^3}) = O(\frac{A^{1.5}}{\epsilon^{1.5}})$, so the gate complexity would be $O(\frac{n^5}{(\epsilon/A)^{1.5}})$.

\subsection*{C.2 Direct PEA (First Order)}
\noindent
To do Direct PEA, we need $n$ qubits to represent the system state and $m=\lceil\log_2(L)\rceil$ qubits to represent the ancilla state. We also need at least 1 qubit for multi-control Toffoli gates in B gate construction\cite{barenco1995elementary}. Towards molecular system of $L = O(n^4)$, so the number of required qubits is $O(n)$.\\

\noindent
To meet properties of B, we can use Householder transformation and set it as:
\begin{equation}
    B = I - \frac{2}{\braket{u|u}}_a\ket{u}\bra{u}_a,
\end{equation}
where $\ket{u}_a = B\ket{0}_a- \ket{0}_a$. The complexity of constructing this gate has been analyzed before\cite{daskin2017direct,ivanov2006engineering,bullock2005asymptotically,urias2015householder}. Since Givens rotation $G_{L-2,L-1}(\theta_{L-1})$ can nullify $B_{0,L-1}$, it can also nullify all $B_{j,L-1}$ for $j\neq L-1$ and update $B_{L-1,L-1}$ to 1 due to $B$'s special form. And $G_{L-2,L-1}^T(\theta_{L-1})$ would nullify all $B_{L-1,j}$ except $B_{L-1,L-1}$. For index smaller than $L-1$ but larger than 1, we can do the same thing. Finally we can choose $G_{1,1}(\theta_{1})$ to update last 4 elements of B and obtain an identity matrix. Thus we have:
\begin{align}
    &G_{1,1}(\theta_1)\prod_{i=2}^{L-1}G_{i-1,i}(\theta_{i}) B\prod_{i=L-1}^{2}G^T_{i-1,i}(\theta_{i}) = I\\
    &B = \prod_{i=L-1}^{2} G^T_{i-1,i}(\theta_i) G^T_{1,1}(\theta_1)\prod_{i=2}^{L-1}G_{i-1,i}(\theta_i)
\end{align}

\noindent
In this way, $B$ gate can be obtained as a product of $2L-3$ Givens rotation matrices. Since each Givens rotation matrix can be achieved by at most $m$ $m$-control Toffoli gates, which would cost $O(m^2)$ standard gates \cite{barenco1995elementary,nielsen2002quantum} each, totally $O(Lm^3) = O(L\log^3n)$ gates are required. For $select(V)$ gate, we need $O((n+m)L)$ standard gates. In this way, $U_r$ requires $O(L\log^3n + (n+m)L) = O(n^5)$ gates. $U_0$ only needs $O(m)$ standard gates, so Q also requires $O(n^5)$ standard gates, which leads the gate complexity of $U_q$ to be $O(Nn^5)$. Since $N = O(\frac{1}{(\epsilon/A)^\frac{1}{2}})$, PEA for $D$ digit accuracy would result in a total of $O(2^DNn^5) = O(\frac{n^5}{(\epsilon/A)^{2.5}})$ standard gates.

\subsection*{C.3 Direct PEA (Second Order)}
\noindent
We need $n$ qubits to represent the system state, $m' =\lceil \log_2(L) \rceil + 2$ qubits to represent the ancilla state. So the number of required qubits is still $O(n)$ as the first order direct PEA.\\

\noindent
When constructing $U_{r2}$, gate $U_r$ takes $O(n^5)$ standard gates, gate $P$ takes $O(L) = O(n^4)$ standard gates, $B_2$, $B_2^\dagger$ and phase gate $e^{-i\frac{\pi}{2}}$ only takes a small constant of standard gates. So the gate complexity of $U_{r2}$ is still $O(n^5)$. $Q_2$ also requires $O(n^5)$ standard gates since $U_0^+$ needs O(m) standard gates. Since $N = O(\frac{1}{(\epsilon/A)^{0.1}})$, PEA for $D$ digit accuracy would result in a toltal of $O(2^D N n^5) = O(\frac{n^5}{(\epsilon/A)^{1.3}})$ standard gates.

\subsection*{C.4 Direct Measurement}
\noindent
The number of required qubits for Direct Measurement Method is the sum of system and ancilla qubits: O(n). Since only one $U_r$ gate is enough, the compexity of the standard gates is $O(n^5)$. Since now the result of measurements is a binomial distribution, to measure the Energy $E$ to accuracy(standard deviation) $\epsilon$, we have to make $X = \frac{E^2}{\epsilon^2}$ measurements.

\subsection*{C.5 Variational Quantum Eigensolver}
\noindent
The number of qubits required for Pairwise VQE is $n$, and the gate complexity is $O(n^2d)$, where $d$ is the number of entangling gate layers. Assume we made $X_i$ measurements for calculating $\langle h_i \rangle$, its accuracy(standard deviation) would be $\epsilon_i = \frac{1}{X_i}$. With $X=\sum_{i=1}^L X_i$ measurements, the accuracy of Hamiltonian would be 
\begin{equation}
\epsilon = \sum_{i=1}^L \frac{a_i}{\sqrt{X_i}} \leq \sqrt{\sum_{i=1}^L a_i^2}\sqrt{\sum_{i=1}^L \frac{1}{X_i}} \leq A \sqrt{\sum_{i=1}^L \frac{1}{X_i}}
\end{equation}
If $X_i = \frac{X}{L}$, we have $\epsilon \leq \frac{AL}{\sqrt{X}}$, then we need $X = \frac{A^2L^2}{\epsilon^2} = \frac{A^2n^8}{\epsilon^2}$ measurements to achieve accuracy $\epsilon$. Considering the number of iterations for optimization, $N_{iter}$, the total number of measurements is $\frac{A^2n^8}{\epsilon^2}N_{iter}$.

\end{document}